\documentstyle[12pt]{article}
\begin{document}
\def\thefootnote{\fnsymbol{footnote}}
\begin{flushright}
KANAZAWA-97-19  \\ 
November, 1997
\end{flushright}
\vspace*{2cm}
\begin{center}
{\LARGE\bf Phenomenological neutrino mass matrix for 
neutrino oscillations and dark matter}\\
\vspace{1 cm}
{\Large  Daijiro Suematsu}
\footnote[3]{e-mail: suematsu@hep.s.kanazawa-u.ac.jp}
\vspace {1cm}\\
{\it Department of Physics, Kanazawa University,\\
        Kanazawa 920-11, Japan}
\end{center}
\vspace{2cm}
A phenomenological neutrino mass matrix is proposed to explain the
solar and atmospheric neutrino deficits and to present the neutrino as a
candidate of hot dark matter in the $3\nu_L+3\nu_R$ framework.
The realization of mixing angles which can explain 
the solar and atmospheric neutrino problems is taken as the first
criterion in this construction.
The differences among neutrino mass eigenvalues are introduced as a 
perturbation.
In this scheme the structure of a charged lepton mass matrix is not
severely constrained by the solar and atmospheric neutrino data.
\newpage
In both particle physics and astrophysics
there are many of indications for massive neutrinos.
The deficiency of solar neutrinos\cite{solar}
and atmospheric neutrinos\cite{atm} has been suggested to be explained by 
$\nu_e$-$\nu_x$ and $\nu_\mu$-$\nu_y$ 
oscillations, respectively.
The predicted neutrino masses and mixings from these observations 
for the solar neutrino problem\cite{mass1}\footnote{
There are also the large mixing solutions.
However, we do not consider these solutions in this
paper.} are \\
\begin{equation}
\Delta m_{\nu_x\nu_e}^2\sim (0.3-1.2)\times 
10^{-5}{\rm eV}^2, \qquad
\sin^22\theta\sim (0.4-1.5)\times 10^{-2},
\end{equation}
and for the atmospheric neutrino problem,\cite{mass2}
\begin{equation}
\Delta m_{\nu_y\nu_\mu}^2\sim (4-6)\times
10^{-3}{\rm eV}^2, \qquad
\sin^22\theta ~{^>_\sim}~ 0.85.
\end{equation}
It was also recently suggested that a cold~ +~ hot dark matter
model agrees well with astrophysical observations if there is 
one neutrino species with $\sim 5$ eV mass.\cite{dark,phkc} 

An interesting feature of these indications is that they require 
wide range mixing angles, in particular, a maximal mixing 
among different neutrinos in addition to hierarchically small mass
eigenvalues.
Although this smallness of masses is thought to be explained by
the seesaw mechanism\cite{seesaw} in general, the hierarchy of masses
and the mixing structure will be
completely dependent on models.

Our aim in this paper is to propose a phenomenological neutrino mass
matrix which can explain both of solar and atmospheric neutrino
deficits. We additionally require neutrinos to be hot
dark matter in a cold ~+~ hot dark matter scenario.
In the study of such a neutrino mass matrix we believe that the
realization of these
mixing angles is the most important clue, and we adopt the following 
strategy.
We first prepare a mass matrix which can realize the
required mixing structure among different neutrinos in the {\it zeroth}
approximation in the seesaw framework. After this, we add the mass 
perturbation needed to induce the hierarchical masses without 
disturbing this mixing structure.

In this direction the author has proposed a neutrino mass matrix
in the $3\nu_L+2\nu_R$ framework.\cite{sue}
In the present study we extend this to the $3\nu_L+3\nu_R$ case. 
The motivation of this extension is based on the
consideration of a relation to the charged lepton mass matrix.
The mixing matrix appearing in neutrino oscillation phenomena is
generally influenced by the charged lepton mass matrix.
In the five neutrinos model this is the case.\cite{sue}
The extension to six neutrinos makes it possible to avoid this
situation, as discussed later.
It is also encouraging for this extension that the $3\nu_L+3\nu_R$ 
framework seems to be natural from the viewpoint of the generation
structure of other quarks and leptons. 

We assume the following 
effective mass terms in the $3\nu_L+3\nu_R$ framework
\begin{equation}
-{\cal L_{\rm mass}}=\sum_{i=2,3,4}m_{i}\bar\psi_{Li}N_{R}
    +\sum_{i=1,5}m_{i}\psi_{Ri}N_{R}+{1 \over 2}MN_{R}N_{R}+{\rm h.c.}
\end{equation}  
and we also assume the mass hierarchy
\begin{equation}
M \gg m_{5} \gg m_{4} \sim m_{3} \gg m_{1} \gg m_{2}. 
\end{equation}
Following the seesaw mechanism, a heavy right-handed neutrino
decouples and a mass matrix for the five light states becomes
\begin{equation}
M_0=M\left( \begin{array}{ccccc}
\mu_1^2 & \mu_1\mu_2 & \mu_1\mu_3 & \mu_1\mu_4 & \mu_1\mu_5 \\
\mu_1\mu_2 & \mu_2^2 & \mu_2\mu_3 & \mu_2\mu_4 & \mu_2\mu_5 \\
\mu_1\mu_3 & \mu_2\mu_3 & \mu_3^2 & \mu_3\mu_4 & \mu_3\mu_5 \\
\mu_1\mu_4 & \mu_2\mu_4 & \mu_3\mu_4 & \mu_4^2 & \mu_4\mu_5 \\
\mu_1\mu_5 & \mu_2\mu_5 & \mu_3\mu_5 & \mu_4\mu_5 & \mu_5^2 \\
\end{array}
\right),
\end{equation} 
where $\mu_i= m_{i}/M (\ll 1)$. As is easily checked, $M_0$ is
diagonalized as $U^{(\nu)}M_0U^{(\nu)T}$ by using the matrix
\begin{equation}
U^{(\nu)}=\left(
\begin{array}{cc}
{\cal O}& \\
  & 1\\
\end{array}\right)
\left(
\begin{array}{ccccc}
{\mu_2 \over \xi_1} & -{\mu_1 \over \xi_1} & 0 & 0 & 0 \\
{\mu_1\mu_3 \over \xi_1\xi_2} & {\mu_2\mu_3 \over \xi_1\xi_2} &
-{\xi_1 \over \xi_2} & 0 & 0 \\
{\mu_1\mu_4 \over \xi_2\xi_3} & {\mu_2\mu_4 \over \xi_2\xi_3} &
{\mu_3\mu_4 \over \xi_2\xi_3} & -{\xi_2 \over \xi_3} & 0 \\
{\mu_1\mu_5 \over \xi_3\xi_4} & {\mu_2\mu_5 \over \xi_3\xi_4} &
{\mu_3\mu_5 \over \xi_3\xi_4} & {\mu_4\mu_5 \over \xi_3\xi_4} & 
-{\xi_3 \over \xi_4} \\
{\mu_1 \over \xi_4} & {\mu_2 \over \xi_4} & {\mu_3 \over \xi_4} &
{\mu_4 \over \xi_4} & {\mu_5 \over \xi_4}
\end{array} \right),
\end{equation}
where $\displaystyle \xi_n^2=\sum_{i=1}^{n+1}\mu_i^2$, and ${\cal O}$
is an arbitrary $4\times 4$ orthogonal matrix.\footnote{This
arbitrariness remains because of the degeneracy of mass eigenvalues.
The author thanks M.~Tanimoto for pointing this out.}

In order to investigate a consistent explanation for
the neutrino mixings (1) and (2), 
we need to identify these five states 
with the physical neutrino states.
In this context the constraint from the standard big bang
nucleosynthesis (BBN)\cite{bbn} is useful.
The BBN predicts that the effective neutrino species 
during the primodial nucleosynthesis should be less than 3.3.
This fact severely constrains the mixing angle $\theta$ and squared mass 
difference $\Delta m^2$ for a sterile neutrino $(\nu_s)$ 
mixing with left-handed active 
neutrinos.\cite{nucl1,ssf}
These constraints rule out the large mixing MSW solution of the solar neutrino
problem due to $\nu_e\rightarrow\nu_s$ and also the explanation of the
atmospheric neutrino problem by $\nu_\mu\rightarrow\nu_s$.
In this study we consider the possibility that 
$\psi_1$ and  $\psi_5$ are right-handed sterile neutrinos and
$\psi_2, \psi_3$ and $\psi_4$ are $\nu_{eL}, \nu_{\mu L}, \nu_{\tau
L}$.\footnote{There is another possibility that $\psi_1$ is
identified with $\nu_{eL}$. However, in that case $\psi_2$ plays no
role and it is reduced to the model considered in Ref. 8).}
The solar and atmospheric neutrino deficits are explained by
the small mixing MSW solution due to $\nu_e\rightarrow\nu_s$ and
the $\nu_\mu$-$\nu_\tau$ oscillation, respectively. 

Here it is useful to remind ourselves that the transition probability 
for $\nu_i \rightarrow
\nu_j$ during the time interval $t$ in the vacuum is expressed as 
\begin{equation}
P_{\nu_i\rightarrow\nu_j}(t)=-4\sum_{k\not= k^\prime}
V_{ik}^{(l)}V_{jk}^{(l)}V_{ik^\prime}^{(l)}V_{jk^\prime}^{(l)}
\sin^2\left( {\Delta m^2_{kk^\prime}t \over 4E}\right),
\end{equation}
where $V^{(l)}$ is the Cabibbo-Kobayashi-Maskawa (CKM) matrix for the 
lepton sector, which can be written as\footnote{
Here $V^{(l)}$ is defined as $\nu=V^{(l)\dagger}\tilde\nu$, where $\tilde\nu$
are the mass eigenstates. The bases $\nu$ are chosen so
that the leptonic charged current and the charged lepton mass matrix
are diagonal.}
$V^{(l)}=U^{(\nu)}U^{(l)\dagger}$ by using the diagonalization matrix 
$U^{(l)}$ of the charged lepton mass matrix: $M_{\rm
diag}=U^{(l)}M^{(l)}U^{(l)\dagger}$.

For the time being, we confine our attention to $U^{(\nu)}$ with 
${\cal O}=1$,
assuming that the charged lepton mass matrix is diagonal.
Under this assumption, taking account of (6) and (7), the desired
mixing angles in (1) and (2) can 
be obtained by setting
\begin{equation}
16 ~{^<_\sim}~ {\mu_1 \over \mu_2} ~{^<_\sim}~ 32, \qquad
0.44 ~{^<_\sim}~ {\mu_3 \over \mu_4}~{^<_\sim}~2.3.
\end{equation}
Although the mixing in the neutrino sector can take a suitable pattern
by imposing this mass hierarchy (8),
the rank of $M_0$ is 1, and the required hierarchical mass 
pattern is not realized. To remedy this situation
and remove the arbitrariness ${\cal O}$ in Eq. (6), we need to add 
a mass perturbation to yield the hierarchical mass eigenvalues
without disturbing the mixing structure $U^{(\nu)}$.
As such a mass perturbation, we consider the simplest example, 
\begin{equation}
M_{\rm per}
\simeq
\left( \begin{array}{ccccc}
A\mu_1^2 & B\mu_1\mu_2 & D\mu_1\mu_3 & E\mu_1\mu_4 & F\mu_1\mu_5\\
B\mu_1\mu_2 & C\mu_2^2 & D\mu_2\mu_3 & E\mu_2\mu_4 & F\mu_2\mu_5\\
D\mu_1\mu_3 & D\mu_2\mu_3 & D\mu_3^2 & E\mu_3\mu_4 & F\mu_3\mu_5\\
E\mu_1\mu_4 & E\mu_2\mu_4 & E\mu_3\mu_4 & E\mu_4^2 & F\mu_4\mu_5 \\
F\mu_1\mu_5 & F\mu_2\mu_5 & F\mu_3\mu_5 &F\mu_4\mu_5 & 2F\mu_4^2 \\
\end{array}
\right),
\end{equation}
where $A \sim F$ are parameters which satisfy
\begin{equation}
{A-D\over B-D}={B-D\over C-D}=-{\mu_2^2\over\mu_1^2},\qquad
{F\over D+E}=-{\mu_4^2 \over \mu_5^2}.
\end{equation}
This matrix is also diagonalized by $U^{(\nu)}$ with ${\cal O}=1$,
and the mass eigenvalues of 
$M_0+ M_{\rm per}$ is obtained as
\begin{equation}
M_1=(C-D)\mu_2^2, \quad M_2=0, \quad M_3=(D-E)\mu_3^2, \quad 
M_4=(D+E)\mu_4^2, \quad M_5=M\mu_5^2 .
\end{equation}
In order to realize the desired masses $M_1\sim 10^{-2.5}$~eV,  
$M_4\sim 10^{-1.2}$~eV and $M_5 ~{^>_\sim}~ 10$~ eV for the neutrino 
deficits and the hot dark matter\footnote{
The heavier right-handed neutrino can be a hot dark matter
candidate. The constraint on $M_5$ comes from the consistency with the
BBN,\cite{ssf,sue} which is somehow larger than one of the usual 
predictions.\cite{dark,phkc}} in a manner consistent with (8),
we must introduce at least three new parameters, $C,~D$ and $E$.\footnote{
For the model defined by (12), mass parameters $C$ and $D\simeq E$ are 
related by 
 $C\sim 10^{3.4}D$ and $D$ should be larger than $M$ by two orders
of magnitude.}
Then we can settle our phenomenological mass matrix by taking, for example,
\begin{equation}
m_1\sim 10^{-1-a}, \quad m_2\sim 10^{-2.4-a}, \quad m_3~{^<_\sim}~ m_4\sim
1, \quad m_5\sim 10^2, \quad M\sim 10^{12},
\end{equation} 
where $a(>)$ is a free parameter, and we use GeV units.\footnote{
There are no strong quantitative
constraints on $\mu_4/\mu_5$ and $\mu_1/\mu_3$ 
as long as these are sufficiently small.}

Next we proceed to the constraint on the charged lepton mass matrix in 
the present framework.
The charged lepton mass matrix is also related to neutrino 
oscillation phenomena through the mixing matrix $V^{(l)}$, as is shown in
Eq. (7).
We use a Fritzsch mass matrix\cite{frit} for the charged lepton sector 
here.  
Using a well-known formula in the diagonalization of the Fritzsch mass 
matrix for $U^{(l)}$, we can obtain the CKM matrix elements 
$V^{(l)}_{ij}$ as
\begin{eqnarray}
&&V^{(l)}_{\nu_e e}\sim {\mu_2\over \mu_1}
-{\mu_1\over \mu_3}\sqrt{m_e\over m_\mu}e^{i\sigma},\quad
V^{(l)}_{\nu_e\mu}\sim 
-{\mu_2 \over\mu_1}\sqrt{m_e \over m_\mu}
 -e^{i\sigma}{\mu_1 \over \mu_3},\quad
V^{(l)}_{\nu_e\tau}\sim 
-e^{i\sigma}{\mu_1 \over \mu_3}\sqrt{m_\mu \over m_\tau},\nonumber \\
&&V^{(l)}_{\nu_\mu e}\sim {1\over\sqrt 2}{\mu_2 \over \mu_3}
+{e^{i\sigma}\over \sqrt 2}\sqrt{m_e \over m_\mu}
+{e^{i\tau} \over \sqrt 2}\sqrt{m_e \over m_\tau},\quad
V^{(l)}_{\nu_\mu \mu}\sim {e^{i\sigma} \over \sqrt 2},\quad
V^{(l)}_{\nu_\mu\tau}\sim -{e^{i\tau} \over \sqrt 2}, \nonumber \\
&&V^{(l)}_{\nu_\tau e}\sim 
{1\over\sqrt 2}{\mu_2 \over \mu_3}
+{e^{i\sigma}\over\sqrt 2}\sqrt{m_e \over m_\mu}
-{e^{i\tau} \over \sqrt 2}\sqrt{m_e \over m_\tau},\quad
V^{(l)}_{\nu_\tau\mu}\sim {e^{i\sigma} \over \sqrt 2},\quad
V^{(l)}_{\nu_\tau\tau}\sim {e^{i\tau} \over \sqrt 2}, 
\end{eqnarray}
where $m_e, m_\mu$ and $m_\tau$ are charged lepton mass eigenvalues.
Here it should be noted that the charged lepton 
mass matrix has only a negligible effect on $V_{\nu_e e}^{(l)}$ 
which is relevant to the $\nu_e$-$\nu_s$ oscillation. 
If we set $\mu_1/ \mu_2$ to appropriate values such as those in Eq. (8), 
the small mixing MSW solution of the solar neutrino problem 
works without dependence on the charged lepton sector.
This is very different from the three\cite{bs,matrix} and 
four\cite{sue} light neutrinos
schemes, where the charged lepton sector crucially affects the
$\nu_e$-$\nu_\mu$ mixing.
In the case of the Fritzsch-type mass matrix, this mixing 
becomes too large to explain the solar neutrino deficit by
$\nu_e\rightarrow\nu_\mu$ unless the
phase is taken to be a suitable value.\cite{bs,matrix}
In the present scheme this situation can be avoided, and the Fritzsch
mass matrix is also applicable to the charged lepton sector without
any assumption on the phases.
This is a direct result of the fact that the solar 
neutrino deficit is explained by the $\nu_e$-$\nu_s$ oscillation due
to the extension to the five light neutrinos.
As long as $U^{(l)}$ is approximately diagonal, like the case of 
Fritzsch mass matrix, our scenario is always applicable, independent
of the details of the charged lepton mass matrix. 

Finally, we briefly comment on possible underlying theories 
which may realize the present scenario.
Such models will not be usual grand unified models, in which 
all the right-handed neutrinos are required to be heavy.  
One promising possibility is a superstring inspired $E_6$ model, in which the 
group theoretical constraints on the Yukawa couplings become very weak. 
Usually it is not easy to obtain small neutrino masses and to induce
neutrino oscillations without bringing other phenomenological
difficulties in the framework.\cite{esix}
However, if we introduce unconventional field assignments 
under suitable conditions in the model, it is possible to show that 
similar structure, at least to the mixing which is discussed here, 
can be realized.
A study along this line can be found in Ref. 16).

In summary we proposed the phenomenological neutrino mass matrix which 
could explain the solar and atmospheric neutrino deficits and offered 
a hot dark matter candidate in the $3\nu_L+3\nu_R$ framework. 
In this construction we took the viewpoint that
the mixing structure was the essential ingredient.
The resulting mass matrix can explain the neutrino oscillation
phenomena without constraining the charged lepton sector.
The relation of this model to a LSND result\cite{lsnd} will be presented 
elsewhere.
\vspace{2mm}

This work is supported
by a Grant-in-Aid for Scientific Research from the Ministry of Education, 
Science and Culture(\#08640362).
 

\end{document}